\documentclass[reprint,superscriptaddress,nofootinbib,nobibnotes,bibnotes,amsmath,amssymb,prd]{revtex4-2}
\usepackage{hyperref}
\usepackage{graphicx}
\usepackage{dcolumn}
\usepackage{bm}
\usepackage{color}
\usepackage{xcolor}

\usepackage[british]{babel}
\usepackage{amsmath}
\usepackage{orcidlink}
\usepackage{mathrsfs}
\usepackage{amsfonts}
\usepackage[normalem]{ulem}
\usepackage{booktabs} 

\usepackage{float}
\usepackage[caption=false]{subfig}
\usepackage{multirow}
\usepackage{xfrac}
\usepackage{listings}
\usepackage{enumitem}
\usepackage[T1]{fontenc}
\usepackage{csquotes}
\usepackage{diagbox}
\usepackage{fontawesome}
\usepackage{url}
\usepackage{setspace}
\usepackage{lineno}

\rightlinenumbers
\definecolor{brightpink}{rgb}{1.0, 0.0, 0.5}

\definecolor{codegreen}{rgb}{0,0.6,0}
\definecolor{codegray}{rgb}{0.5,0.5,0.5}
\definecolor{codepurple}{rgb}{0.58,0,0.82}
\definecolor{backcolour}{rgb}{0.95,0.95,0.92}
\definecolor{ao}{rgb}{0.0, 0.5, 0.0}
\lstdefinestyle{mystyle}{
    backgroundcolor=\color{backcolour},   
    commentstyle=\color{codegreen},
    keywordstyle=\color{magenta},
    numberstyle=\tiny\color{codegray},
    stringstyle=\color{codepurple},
    basicstyle=\ttfamily\footnotesize,
    breakatwhitespace=false,, 
    breaklines=true,,       
    captionpos=b,,, 
    keepspaces=true,,       
    numbers=left,,, 
    numbersep=5pt,,
    showspaces=false,,      
    showstringspaces=false,
    showtabs=false,,
    tabsize=2
}
\makeatother
\newcommand{\iu}{\mathrm{i}}
\newcommand{\ellmax}[0]{\ell_{\rm max}}
\newcommand{\bigO}[0]{\mathcal{O}}

\newcommand{\cunusht}[0]{\texttt{cunuSHT}}
\newcommand{\ducc}[0]{$\texttt{DUCC}$}
\newcommand{\shtns}[0]{$\texttt{SHTns}$}

\newcommand{\lenspyx}[0]{$\texttt{lenspyx}$}

\newcommand{\cufinufft}[0]{$\texttt{cufinuFFT}$}
\newcommand{\jax}[0]{$\texttt{JAX}$}

\newcommand{\rsht}[0]{rSHT}

\newcommand{\GPUdesc}[0]{A-100}
\newcommand{\GPUmem}[0]{$80$}

\newcommand{\CPUdesc}[0]{Intel Xeon Gold 8358 Processor}
\newcommand{\nthreads}[0]{$32$}
\newcommand{\suf}[0]{$5$} 

\begin{document}
\preprint{}
\title{\cunusht{}: GPU Accelerated Spherical Harmonic Transforms on Arbitrary Pixelizations}

\newcommand{\CCA}{\small Center for Computational Astrophysics, Flatiron Institute, 162 5th Avenue, New York, NY 10010, USA}
\newcommand{\Geneve}{Universit\'e de Gen\`eve, D\'epartement de Physique Th\'eorique et CAP, 24 Quai Ansermet, CH-1211 Gen\`eve 4, Switzerland}
\newcommand{\CNRS}{\small ISTerre, Université de Grenoble 1, CNRS, F-38041 Grenoble, France}
\newcommand{\MPA}{\small Max-Planck Institut für Astrophysik, Karl-Schwarzschild-Str. 1, 85748 Garching, Germany}
\author{Sebastian Belkner\orcidlink{https://orcid.org/0000-0003-2337-2926}}
\affiliation{\CCA}
\affiliation{\Geneve}
\author{Adriaan J.\ Duivenvoorden \orcidlink{https://orcid.org/0000-0003-2856-2382}}
\affiliation{\CCA}
\author{Julien Carron\orcidlink{https://orcid.org/0000-0002-5751-1392}}
\affiliation{\Geneve}
\author{Nathanael Schaeffer \orcidlink{https://orcid.org/0000-0001-5206-3394}}
\affiliation{\CNRS}
\author{Martin Reinecke}
\affiliation{\MPA}

{}
\date{\today}

\begin{abstract}
We present \cunusht{} \faGithub, a general-purpose Python package that wraps a highly efficient CUDA implementation of the nonuniform spin-$0$ spherical harmonic transform. The method is applicable to arbitrary pixelization schemes, including schemes constructed from equally-spaced iso-latitude rings as well as completely nonuniform ones. The algorithm has an asymptotic scaling of $\bigO{(\ellmax^3)}$ for maximum multipole $\ellmax$ and achieves machine precision accuracy. While \cunusht{} is developed for applications in cosmology in mind, it is applicable to various other interpolation problems on the sphere. We outperform the fastest available CPU algorithm by a factor of up to \suf{} for problems with a nonuniform pixelization and $\ellmax>4\cdot10^3$ when comparing a single modern GPU to a modern \nthreads{}-core CPU. This performance is achieved by utilizing the double Fourier sphere method in combination with the nonuniform fast Fourier transform and by avoiding transfers between the host and device. For scenarios without GPU availability, \cunusht{}  wraps existing CPU libraries. \cunusht{} is publicly available and includes tests, documentation, and demonstrations.
\end{abstract}

\maketitle
\textbf{keywords:} Nonuniform Spherical Harmonic Transform, Nonuniform Fast Fourier transform, Cosmic Microwave Background Weak Lensing, CUDA

\section{Introduction}
Spherical harmonic transforms (SHTs) are a key ingredient in signal processing for data sets on the 2-sphere. They  are extensively used in active research fields such as cosmology (both in studying the cosmic microwave background (CMB) \cite{hivon_2002,Planck:2018lbu, SPTpol:2020rqg, BICEP2:2018kqh, ACT:2023dou, NASAPICO:2019thw} and large-scale structure of the Universe \cite{scharf_1993,hikage_2011}), gravitational waves \cite{spectrecode}, meteorology \cite{Nils:2013}, solar physics \cite{Brun:2009}, or solving partial differential equations on the sphere \cite{Browning:1989}.

Modern data sets routinely require the evaluation of the SHT up to $\ellmax = \bigO(10^4)$ for maps with $N\sim\ellmax^2$ pixels. Here $\ellmax$ denotes the largest multipole $\ell$ considered in the SHT. 
A direct evaluation of the SHT scales as $\bigO(\ellmax^4)$, which is intractable for large problems. Reducing the computational complexity is thus crucial. One standard optimization is achieved by pixelizing the sphere into rings of constant latitude with equi-angular spaced samples, reducing the problem to $\mathcal{O}(\ellmax^3)$.
We will refer to this setup as the ring spherical harmonic transform (rSHT).\footnote{There exist algorithms for the rSHT that asymptotically scale as $\bigO{(\ellmax^2\log^2(\ellmax))}$ \cite{DRISCOLL:1994202,pott:1996,pott:1998,KeKuPo:2009,Seljebotn:2012} or $\bigO{(\ellmax^2\log^2(\ellmax)/\log\log(\ellmax)})$~\cite{Hale:2015}. While there have been improvements over the last few years \cite{slevinsky2017fast,Yin:2019} such methods require high memory usage and significant pre-computations. A general purpose implementation that is competitive with the $\bigO(\ellmax^3)$ counterpart has yet to be published.}
See \cite{Reinecke:2013,Schaeffer2012EfficientSH} for modern implementations of the rSHT. 
For cases where the transform has to be evaluated on irregularly sampled pixels, ring sampling is not possible. We will refer to this more general setup as the nonuniform spherical harmonic transform (nuSHT). Notable applications of the nuSHT are CMB weak lensing \cite{Lewis:2006fu}, ray tracing \cite{Fabbian:2018,ferlito2024raytracing}, or fields where the pixelization changes over time. 
It should be noted that due to their computational complexity, the rSHT and nuSHT often become the bottleneck in iterative algorithms that repeatedly apply the transforms~\cite{wandelt_2004,Carron:2017mqf}.

In the field of cosmology, nuSHTs are routinely solved via an \rsht{} and subsequent interpolation to the nonuniform points using bicubic splines~\cite{Lewis:2005tp} or a Taylor series expansion~\cite{Naess:2013}. These methods scale as $\bigO{(\ellmax^3})$ but only reach relatively low accuracy.
A different $\bigO{(\ellmax^3})$ nuSHT method, proposed by~\cite{basak:2008}, makes use of the double Fourier sphere (DFS) method~\cite{Merilees:1973} in combination with the nonuniform fast Fourier transform (nuFFT) to achieve an accurate nuSHT algorithm. Several other implementations of this setup exist~\cite{townsend:2016computing,KeKuPo:2009}. 
Recently, the \lenspyx{} and \ducc{} libraries \cite{Reinecke:2013,Reinecke2023ImprovedCM} implemented a highly efficient and machine precision accurate implementation of the nuSHT based on the DFS method. 

In recent years, there has been a dramatic increase in the  availability and usage of graphics processing units (GPUs) dedicated to scientific computing. 
This development drives the need for GPU based codes and provides an opportunity for increased performance of existing methods.
GPUs are optimized for single-instruction multiple-data (SIMD) applications and provide multi-threading well beyond what is achievable with CPUs. Highly parallelizable algorithms can thus greatly benefit from the GPU architecture. In principle, the evaluation of the SHT allows for a large amount of parallel computation, making it a natural target for a GPU implementation. The work presented in~\cite{Hupca_2012,szydlarski:2013parallel} was one of the first that explored the use of GPUs for SHTs in the context of cosmology. 

Robust and efficient \rsht{} GPU implementations have been developed in recent years. Notable examples are \shtns{} \cite{Schaeffer2012EfficientSH}, supporting the spin-$0$ and $1$ rSHTs, and \texttt{S2HAT}\footnote{\url{https://apc.u-paris.fr/APC_CS/Recherche/Adamis/MIDAS09/software/s2hat/s2hat.html}} \cite{szydlarski:2013parallel}, which supports spin-$n$ transforms. \texttt{S2FFT} \cite{Price:2024} is a recent \jax{} implementation of the spin-$n$ rSHT that provides differentiable transforms. A first implementation of the nuSHT in a cosmological context on GPUs was presented in \cite{lizancos2024harmonic}.

We present \cunusht{}, a CUDA accelerated nuSHT algorithm on the GPU.
This is, to our knowledge, the first publicly available nuSHT GPU algorithm that reaches machine precision accuracy and achieves significant speed-up compared to the fastest CPU algorithms. We achieve this by carefully combining existing robust and efficient GPU implementations of the \rsht{} and nuFFT algorithm. \cunusht{} does not require memory allocation or calculation on the host, which allows it to be incorporated in GPU-based software. 

The remainder of the paper is organized as follows. In Section~\ref{sec:nuSHT} we introduce notation and definitions, and present in qualitative terms our implementation. Section~\ref{sec:implementation} discusses the implementation on the GPU. Section~\ref{sec:benchmark} shows benchmarks and results. We conclude in Section~\ref{sec:conclusion}. A series of appendices collects further details.

\section{Nonuniform Spherical Harmonic transform}\label{sec:nuSHT}
We introduce our notation and conventions in ~\ref{subsec:def}, and define the spherical harmonic transform operations that we implement in this paper. We describe the double Fourier sphere method in~\ref{subsec:impl}.

\subsection{Definition and properties}\label{subsec:def}
The (spin-0) spherical harmonic functions $Y_{\ell}^{m}(\theta, \phi)$, with quantum numbers $\ell$ and $m$, with $-\ell \leq m \leq \ell$, are given by,
\begin{equation}\label{eq:Ylm}
Y_{\ell}^{m}(\theta, \phi) = P_\ell^m(\theta)e^{\iu m\phi},
\end{equation}
where $P_\ell^m(\theta)$ are the associated Legendre polynomials.
A ``general'' (or ``nonuniform'') spherical harmonic transform (nuSHT) is a linear transformation between a set of spherical harmonic coefficients and field values defined at arbitrary locations on the sphere.
We distinguish two types of transforms, with nomenclature inspired by nonuniform Fourier transform literature\footnote{The \ducc~package uses the names \texttt{adjoint}$\_$\texttt{synthesis}$\_$\texttt{general} and \texttt{synthesis}$\_$\texttt{general} for type 1 and type 2.}~\cite{Barnett2018APN}:
\begin{itemize}
\item \emph{Type 1} (also ``adjoint nuSHT'', the adjoint operation to type 2 below): 
     given as input a set of $N$ values $f_i$, and $N$ locations $(\theta_i, \phi_i)$, desired are the coefficients $c_{\ell m}$ defined by,
    \begin{equation}\label{eq:adjointSHTbase}
        c_{\ell m} = \sum_{i = 1}^N f_i \:Y_{\ell}^{\dagger m}(\theta_i, \phi_i)\,,
    \end{equation}
    for $\ell$ up to some band-limit $\ellmax$. We want the result to match a target accuracy $\epsilon$ requested by the user.

\item \emph{Type 2}: given as input a set of harmonic coefficients $c_{\ell m}$ up to some band-limit $\ellmax$, and a set of $N$ locations $(\theta_i, \phi_i)$, desired are the field values,
\begin{equation}\label{eq:SHTbase}
f_i = \sum_{\ell=0}^{\ellmax} \sum_{m=-\ell}^{\ell} c_{\ell m} Y^m_{\ell}(\theta_i, \phi_i)\,,
\end{equation}
again respecting a target accuracy $\epsilon$ as requested by the user.
\end{itemize}
In matrix notation, type 2 may be written as,
\begin{equation}
    \mathbf{f} = \mathbf{Y} \mathbf{c},
\end{equation}
where the vector $\mathbf{c}$ collects the harmonic coefficients, the vector $\mathbf{f}$ the output field values, and the entries of the matrix $\mathbf{Y}$ are the spherical harmonics. Type 1 is represented by the adjoint matrix $\mathbf{Y}^\dagger = \left[\mathbf{Y}^t\right]^*$. It is worth noting that type 1 is not the inverse to type 2, except in special cases. We also use the qualifiers type 1 and type 2 for the analogous nonuniform (or uniform) Fourier transforms, where the spherical harmonics and coefficients are replaced by their plane wave counterparts.

In typical applications, the total number of points $N$ is comparable to the squared band-limit, $\ellmax^2$. In this case the naive computational complexity of these operations is $\bigO(\ellmax^4)$.

\subsection{Double Fourier sphere method}\label{subsec:impl}
We adopt the approach proposed in~\cite{Reinecke2023ImprovedCM} and implement type 1 and type 2 transforms using the double Fourier sphere (DFS) method. 
In this approach, the matrix $\mathbf{Y}$ of the type 2 nuSHT is decomposed into 4 matrices,
\begin{equation}\label{eq:SHTfast}
\mathbf{Y} = \mathbf{NFDS}\,. 
\end{equation}
The final operation $\mathbf{N}$ is a nonuniform Fourier transform of type 2 to the given locations, and the role of $\mathbf{FDS}$ is to produce the needed Fourier coefficients.

The matrix $\mathbf{S}$ is an iso-latitude \rsht{}, that transforms the input harmonic coefficients onto an equi-spaced grid in both $\phi$ and $\theta$, covering the entire sphere.
\begin{figure}
    \centering
    \includegraphics[width=0.29\textwidth]{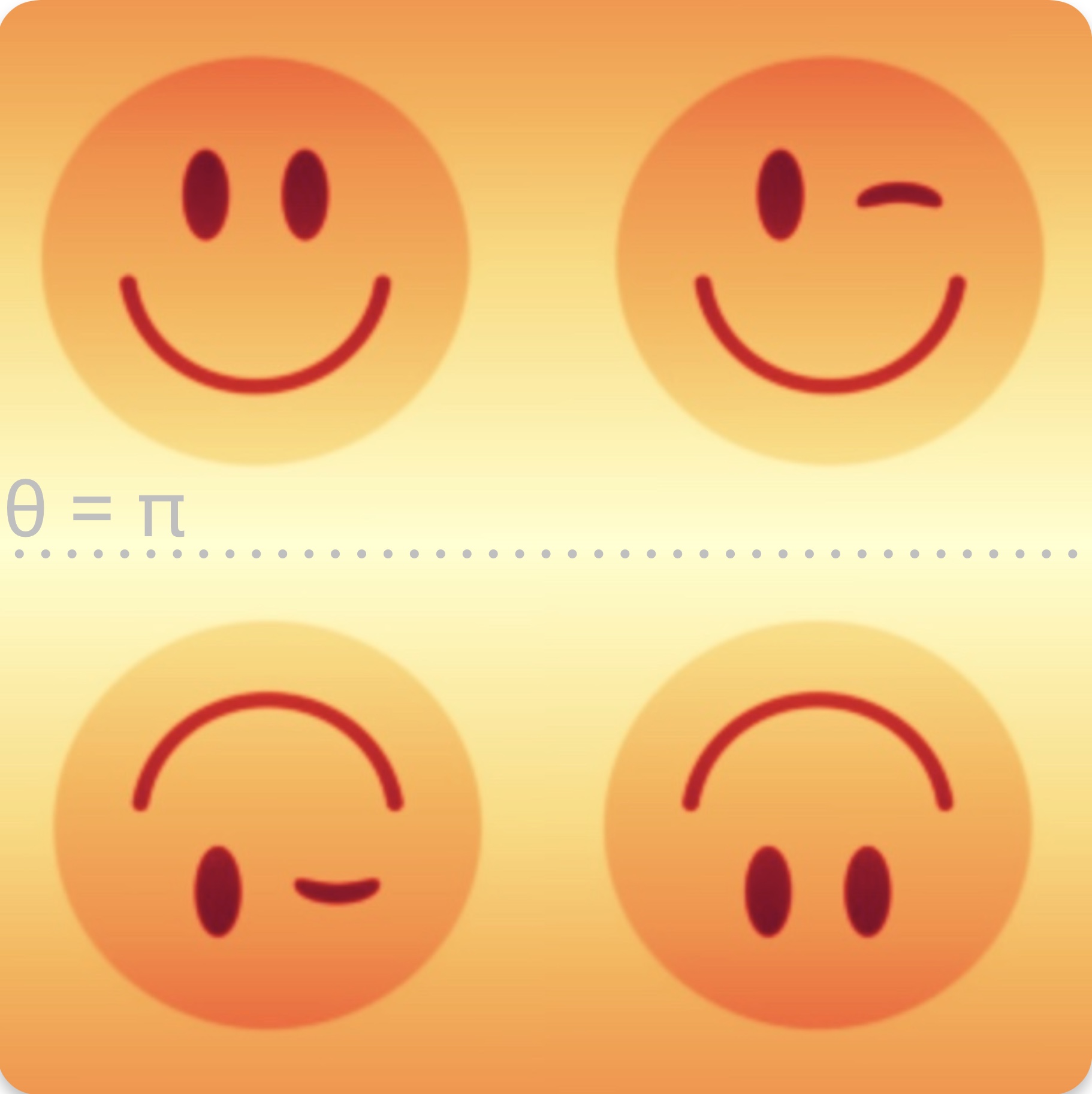}
    \caption{Illustration of the doubling step in the DFS method. The upper half shows a spherical map calculated on a rectangular grid and is mirrored along the $\theta = \pi$ axis. The mirrored image is split in half across the $\phi$-direction in the center, and swapped. The result is a function on the torus, with a 2D ordinary Fourier series having the exact same band-limit as the spherical harmonic series of the original map defined on the sphere. This allows the use of efficient nonuniform Fast Fourier Transform for accurate interpolation.}
    \label{fig:doubling}
\end{figure}
$\mathbf{D}$ is a ``doubling'' operation, that extends the range of $\theta$ from $[0, \pi]$ to $[0, 2\pi)$, see Fig.~\ref{fig:doubling}. The doubling is performed by extending the meridians across the south pole back up to the north pole. The essential point is that the resulting map, seen as a map on the doubly-periodic torus, has a standard Fourier series with exactly the same Fourier band-limit\footnote{This may be seen for example from the well-known Fourier representation of the Wigner $d$-matrices~\cite{Risbo96, Huffenberger:2010hh}, and using the relation $P^m_\ell(\theta) = d^\ell_{m0}(\theta) \sqrt{ (2\ell + 1)/4\pi}$.} as the spherical harmonic band-limit of the input array $c_{\ell m}$.
Finally, $\mathbf{F}$ is simply the standard 2D Fourier transform that produces the Fourier coefficients input to $\mathbf{N}$ from the doubled map.

The adjoint operator $\mathbf{Y}^{\dagger}$, or the type 1 nuSHT is, by definition,
\begin{equation}\label{eq:adjointSHTfast}
\mathbf{Y^{\dagger}} = \mathbf{S^{\dagger} D^{\dagger} F^{\dagger} N^{\dagger}}\,.
\end{equation}
$\mathbf{N}^{\dagger}$ is a nuFFT of type 1 that produces Fourier frequencies from the input locations and field values. $\mathbf{F}^\dagger$ produces from these frequencies a 2D map on the torus. $\mathbf{D}^{\dagger}$ (the adjoint doubling matrix) effectively ``folds'' this doubled map. The resulting map is $2\pi$-periodic in the $\phi$-direction, and the $\theta$-direction goes again from $0$ to $\pi$. The map is then transformed to harmonic coefficients with $\mathbf{S}^{\dagger}$, an iso-latitude type 1 rSHT.

\section{Implementation}\label{sec:implementation} 
We discuss the concrete  GPU implementation. Readers interested in the CPU equivalent may consult \cite{Reinecke2023ImprovedCM}.

A GPU is designed to efficiently apply a single instruction on multiple data (SIMD). On the hardware side, it achieves this with Streaming Multiprocessors (SMs) (at the order of 100), that contain a number of simple processors for arithmetic operations (at the order of 100) that execute ``warps'' of 32 threads in parallel.

On the software side, a GPU accelerated program is executed via a number of threads that are arranged in thread blocks. The GPU is responsible for distributing the thread blocks across the SMs.  High throughput is achieved by overloading SMs with many threads as to hide data latency and by ensuring that memory is accessed in multiples of the warp size.

We differentiate between the GPU memory that is ``close'' to the processor units and can be accessed fast by the device, and host memory, that is managed by the host system of the GPU, and which is generally slow to access by the GPU. We show data transfer benchmarks in Appendix~\ref{ap:transfer}. Our implementation avoids data transfer and usage of host memory altogether; intermediate results are kept in GPU memory. This is realized by \texttt{cupy}-arrays in combination with a \texttt{C++}-binding \texttt{nanobind}, \cite{nanobind}, handily providing a \texttt{nanobind}-\texttt{cupy} interface.

Our implementation of the individual operators $\mathbf{N}$, $\mathbf{D}$, $\mathbf{F}$, and $\mathbf{S}$ and their adjoints are realised as follows.

For the (adjoint) synthesis  ($\mathbf{S}^{\dagger}$) $\mathbf{S}$, we use the highly efficient software package \shtns{} \cite{Schaeffer2012EfficientSH}, and calculate the SHTs onto a Clenshaw-Curtis (CC) grid. The GPU implementation requires the sample size to be divisible by~$4$.

For iso-latitude rings, in order to achieve best efficiency for the Legendre transform part (the $\theta$ part of the transform, which is the critical part), modern top-performing CPU and GPU codes like \shtns{}  use on-the-fly calculation of $P_\ell^m(\theta)$ using efficient
recurrence formulas put forward recently~\cite{KeiichiISHIOKA20182018-019}.
This allows to keep memory usage low: indeed, only the recurrence coefficients that are independent of $\theta$ need to be stored, which requires only $\bigO(\ellmax^2)$ memory, the same order as the data.
It leaves two dimensions along which to parallelize: $\theta$ and $m$, and requires a sequential loop over $\ell$ to compute the $P_\ell^m(\theta)$ recursively.
When $\ellmax$ is larger than 500 to 1000, this leaves enough parallelization opportunities to efficiently use all of the GPU compute units. The computational complexity stays $\bigO(\ellmax^3)$.

The (adjoint) doubling ($\mathbf{D}^\dagger$) $\mathbf{D}$ is implemented via CUDA, and we write the arrays in a $\theta$-contiguous memory layout, as required by SHTns to keep high efficiency for the Legendre transform. The computational complexity is $\bigO(\ellmax^2)$.

For the type-$1$ and type-$2$ nuFFT in $2$-dimensions ($\mathbf{N}^\dagger$, $\mathbf{N}$), we use \cufinufft{} \cite{Shih2021cuFINUFFTAL} in double precision. 
The nuFFT method works by utilizing the Fourier transform convolution theorem, and interpolation or convolution onto a slightly larger, up-sampled grid. Highly accurate versions use kernels whose error $\epsilon$ decrease exponentially as a function of the up-sampling factor.
The computational complexity (without planning phase) is $\bigO(\ellmax^2\log(\ellmax) + \ellmax^2|\log^2(\epsilon)|)$ in 2 dimensions.
It is worth mentioning that we use the guru interface to \cufinufft{} to initialize the nuFFT plans. The plans allow for repeated and fast transforms, without re-initialization. However, this planning step is the most time consuming operation and is therefore done before calling the functions.

For the FFT operations $\mathbf{F}$ (type 2) and $\mathbf{F}^\dagger$ (type 1), we use the package \texttt{cupyx} \cite{cupy_learningsys2017} and its \texttt{cuFFT}\footnote{\url{https://developer.nvidia.com/cufft}} integration therein. For the type 1 Fourier synthesis, we use double precision accuracy. For the type 2 Fourier synthesis, we use single precision accuracy for $\epsilon\leq10^{-6}$, which increases speed at the cost of a negligible decrease in effective accuracy on the final result. The computational complexity is $\bigO(\ellmax^2\log(\ell_{\rm max}))$.

FFTs become particularly fast if the prime factorization for the sample size gives many small prime numbers, in the following referred to as a \emph{good number}. If additional constraints are put on the sample size, good numbers may be more difficult to find, see Appendix \ref{ap:goodnumber} for a discussion and concrete definition.

Ignoring the scaling with accuracy, the overall asymptotic computational complexity (for both type 1 and 2) is, 
\begin{align}\label{eq:ccx}
\begin{split}\bigO(\ellmax^3) + \bigO(\ellmax^2\log(\ellmax)) + \bigO(\ellmax^2)\end{split}\,.
\end{align}
Here, we assume that the number of uniform and nonuniform points are about the same.

$\mathbf{Y}$ (and $\mathbf{Y}^\dagger$) could be further optimized: $\mathbf{S}$ and $\mathbf{F}$ both contain a Fourier transform in $\phi$-direction, effectively cancelling out each other. Avoiding this reduces $\mathbf{F}$ to a $1$-dimensional Fourier transform and $\mathbf{S}$ to a Legendre transformation. This optimization is implemented in the CPU implementation in \ducc{}. We leave this optimization to a future study for the GPU implementation.  We only expect a large speed up for the $\mathbf{F}$ operator in \cunusht{}, which takes about 10 to 20\% of the total runtime.

For CMB weak lensing applications, we also provide a pointing routine implemented via \texttt{CUDA}, see Appendix \ref{ap:pointing}.

\section{Benchmark}\label{sec:benchmark} 
We present the scaling and execution time of our implementation. We take as a use case an application in CMB weak lensing~\cite{Lewis:2006fu}. 
CMB weak lensing describes the deflection of primordial CMB photons by mass fluctuations along the line of sight as they travel through the Universe. These small deflections (their root-mean-square is of the order of a few arcminutes) are large enough to be detectable in CMB sky maps~\cite{Aghanim:2018oex, ACT:2023dou, SPT:2023jql}. 
For some applications, it is necessary to simulate these deflections accurately and efficiently~\cite{Hirata:2003ka, Carron:2017mqf, Belkner:2023duz}.

\begin{figure*}
    \centering
        \includegraphics[width=0.99\textwidth]{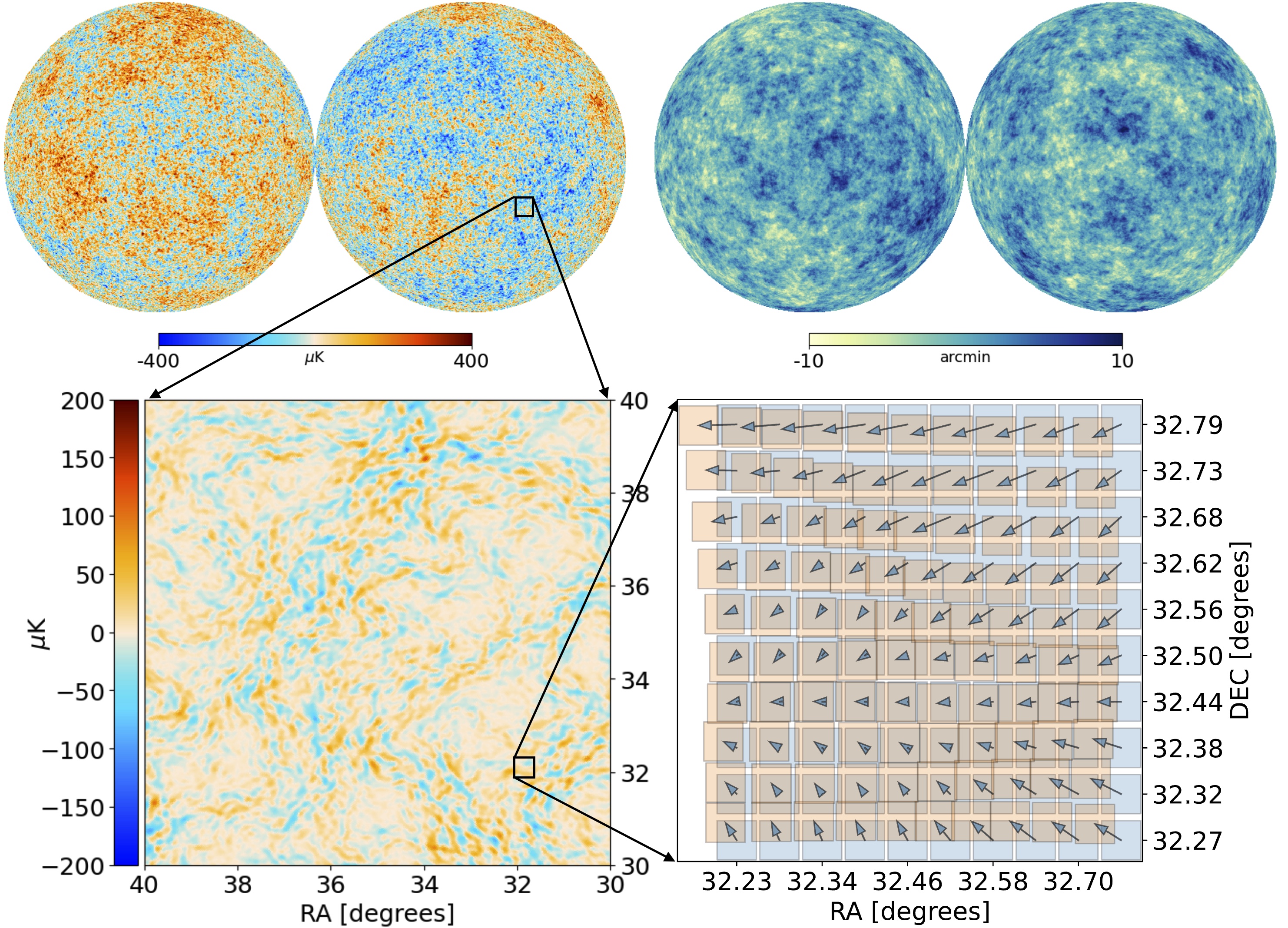}
    \caption{Setup and simulated data, here for a problem size of $\ellmax=3095$, $N\approx 2\cdot 10^7$. The top left (right) panel shows a typical CMB temperature map (a typical deflection field) in orthographic projection. The bottom left plot shows a $10\times 10$ degree detail view of the difference between the undeflected and deflected CMB in Cartesian projection. On the bottom right, we show a $0.5\times 0.5$ degree detail view comparison between the uniform grid (blue squares), nonuniform grid (orange squares), and the relation between them as indicated by the black arrows.}
    \label{fig:benchmark_setup}
\end{figure*}
Owing to the deflections, the CMB intensity field $\tilde f(\theta, \phi)$ observed at location $(\theta, \phi)$ is the un-deflected field $f$ at another location,
\begin{equation}\label{eq:cmbsky}
    \tilde{f}(\mathbf{\theta},\mathbf{\phi}) = f(\mathbf{\theta'},\mathbf{\phi'})\,,
\end{equation}
where $(\theta', \phi')$ depends on $(\theta, \phi)$ in a smooth way. Details on how these angles relate to each other are given in  Appendix \ref{ap:pointing}. For the type-2 nuSHT, the inputs are the set of angles $(\theta', \phi')$ and the spherical harmonic coefficients of the un-deflected field $f$. For type-1, the input are the same set of angles and a set of values of $\tilde f$.
The set up of our use case is shown in Fig.~\ref{fig:benchmark_setup}. The top left panel shows the input CMB map, the deflection field is shown on the top right. Both are shown in orthographic projection. The bottom maps show a detail view of $10\times 10$ degree of the difference between the input and deflected map (left panel) and a detail view of approximately $0.5\times 0.5$ degree (right panel) of the nonuniform point (orange squares) relative to the uniform point locations (blue squares). We use the same setup for the adjoint operation, in which case the deflected map becomes the input, and the result becomes the adjoint SHT coefficients.

Benchmarks are run on an NVIDIA \GPUdesc{} GPU with \GPUmem{} GB of memory and an \CPUdesc{} with 32 cores. We set the number of threads to \nthreads{} for the CPU benchmarks.
If not stated otherwise, we choose the following parameters (that mostly affect the nuFFT): an up-sampling factor of $1.25$ to reduce the memory usage at a small price of increased computation time\footnote{This somewhat low up-sampling factor reduces the minimal possible accuracy, in double precision, to $10^{-10}$ due to its dependence on the size of the up-sampled grid, but can easily be changed by the user, if needed.}, a \texttt{gpu\_method} utilizing the hybrid scheme, called \emph{shared memory}, and the default kernel evaluation method. The resulting map (or input map in the case of the adjoint) is calculated on a Gauss-Legendre grid.

With this implementation, we can solve problem sizes of up to $\ellmax\sim9000$ on an \GPUdesc{} with \GPUmem{} GB, by using the pre-computed nuFFT plans and keeping all necessary intermediate results in memory.

\begin{figure*}
    \centering
    \includegraphics[width=.45\textwidth]{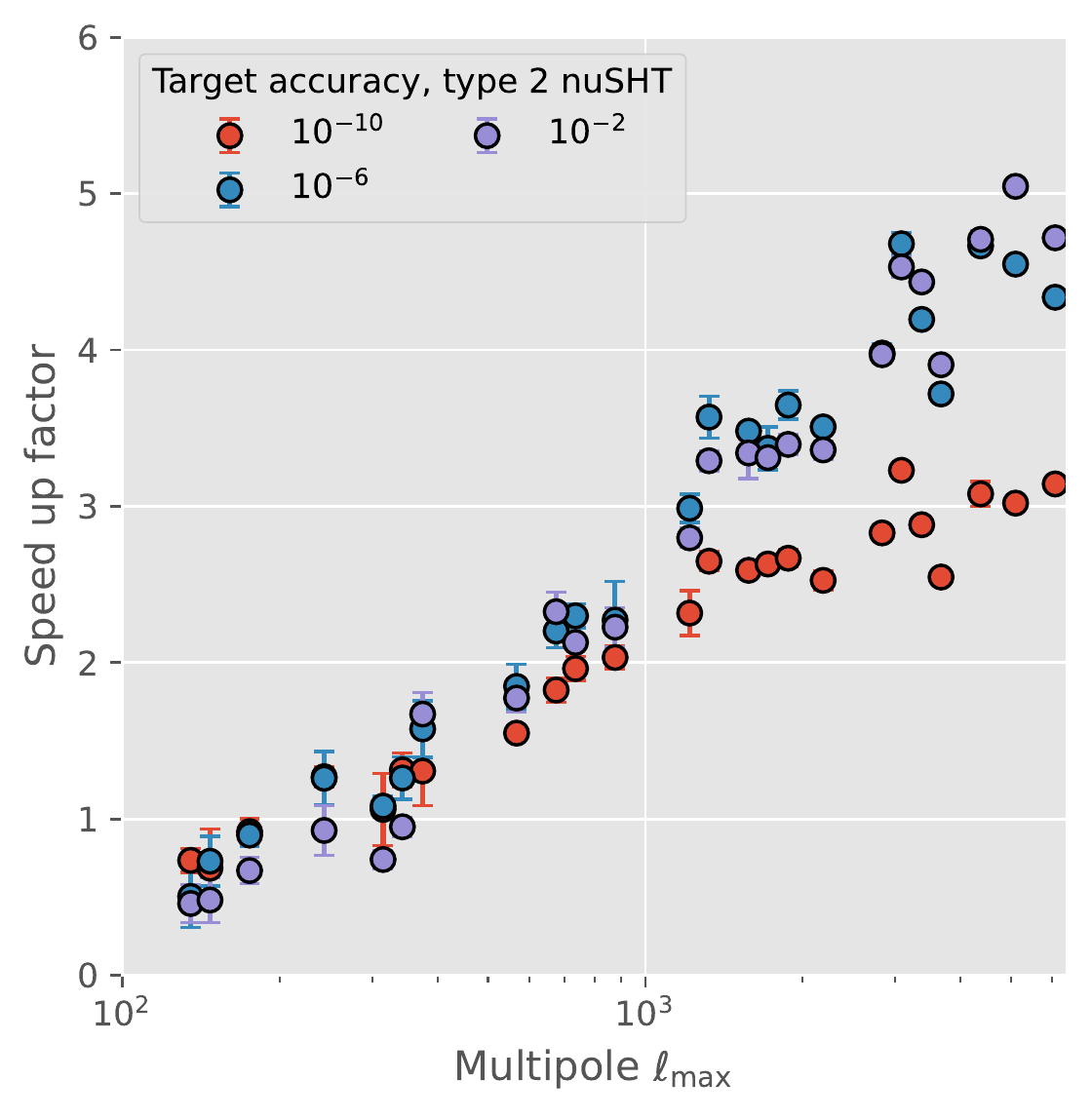}
    \includegraphics[width=.45\textwidth]{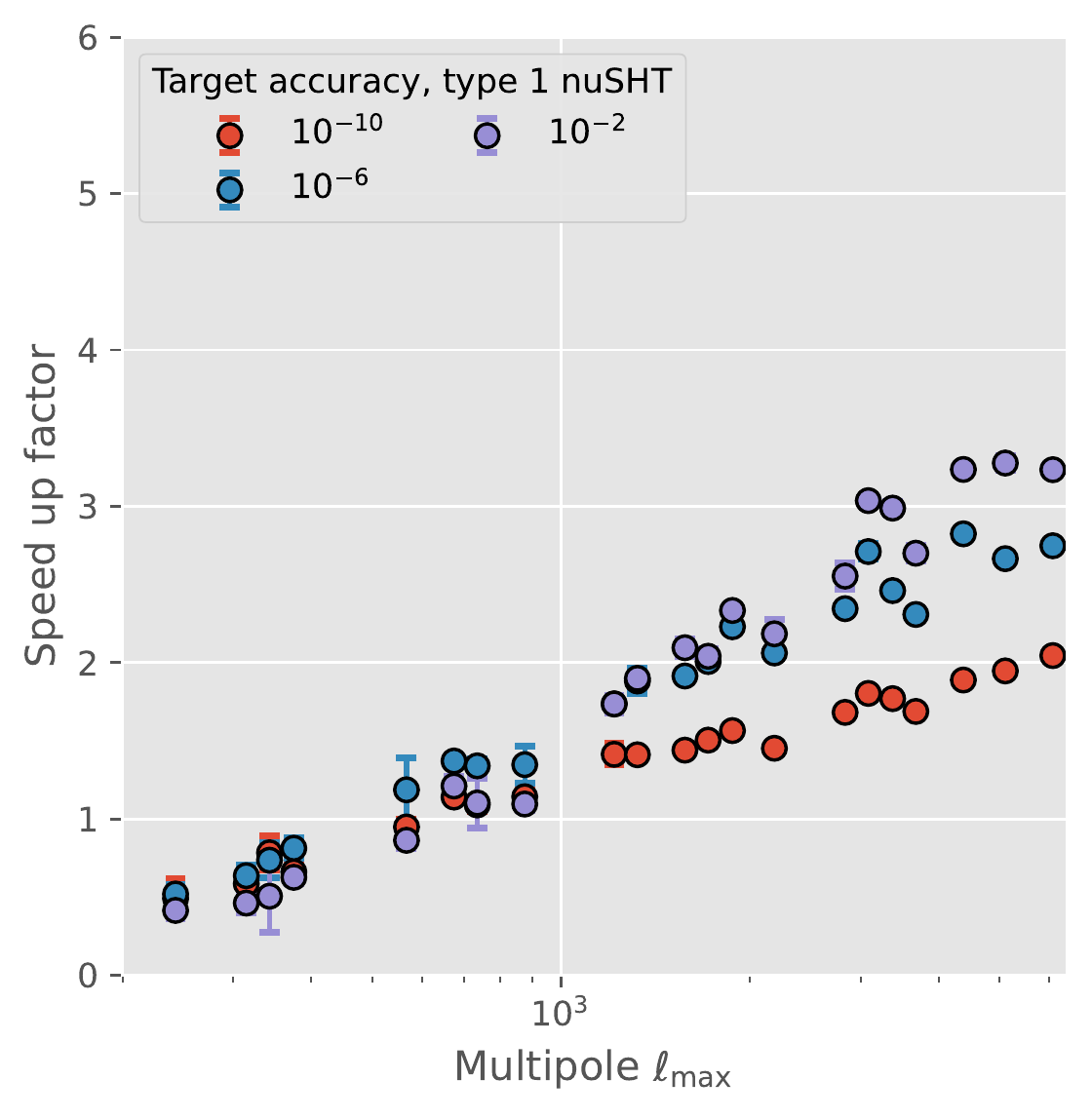}
    \caption{\cunusht{} execution time comparison against \ducc{} for type 2 (type 1) nuSHT shown in the left (right) panel, comparing one \GPUdesc{} against one \CPUdesc{} with 32 cores. A speed up factor $>1$ means that the GPU is faster, and we show the result for different target accuracies. The error bars show the $\pm1\sigma$ variance calculated from $5$ runs. The GPU takes over at $\ellmax\sim 3\cdot10^2$ for type 2, and between $\ellmax\sim 4\cdot10^2$ for type 1. The speed up increases up to \suf{} (3) for large $\ellmax$ left (right) panel.}
    \label{fig:speedup}
\end{figure*}
Fig.~\ref{fig:speedup} shows the speed up of the GPU algorithm compared to the CPU as a function of $\ellmax$  for different accuracies. The left panel shows the evaluation of Eq.~\eqref{eq:SHTfast}, the right panel shows Eq.~\eqref{eq:adjointSHTfast}. The $\pm1\sigma$ variance from $5$ runs is indicated by error bars.

For type 2 nuSHT, we reach a speed up between 1 and 5 times for single precision and 1 to 3 times for double precision, with the speed up increasing with increasing $\ellmax$. This increase is expected due to better parallelizability for GPUs for higher $\ellmax$. The speed up is mostly independent of the target accuracy, with the smallest accuracies tending to perform better.
For type 1 nuSHT, we find lower speed up factors, which may indicate further improvements. However, type 1 directions are generally expected to perform worse on the GPU due to the shear amount of threads that have to be written concurrently to the same memory location.
Nevertheless, the GPU code either performs almost as good as the CPU code (small $\ellmax$), or better by a factor of up to $3$ for large problem sizes. For high $\ellmax$, the speed up depends on the accuracy, with lower accuracies performing better. For double precision ($\epsilon=10^{-10}$), the speed up is diminished due to the double precision penalty that we pay in our implementation.

\begin{figure*}
    \centering
    \includegraphics[width=0.49\textwidth]{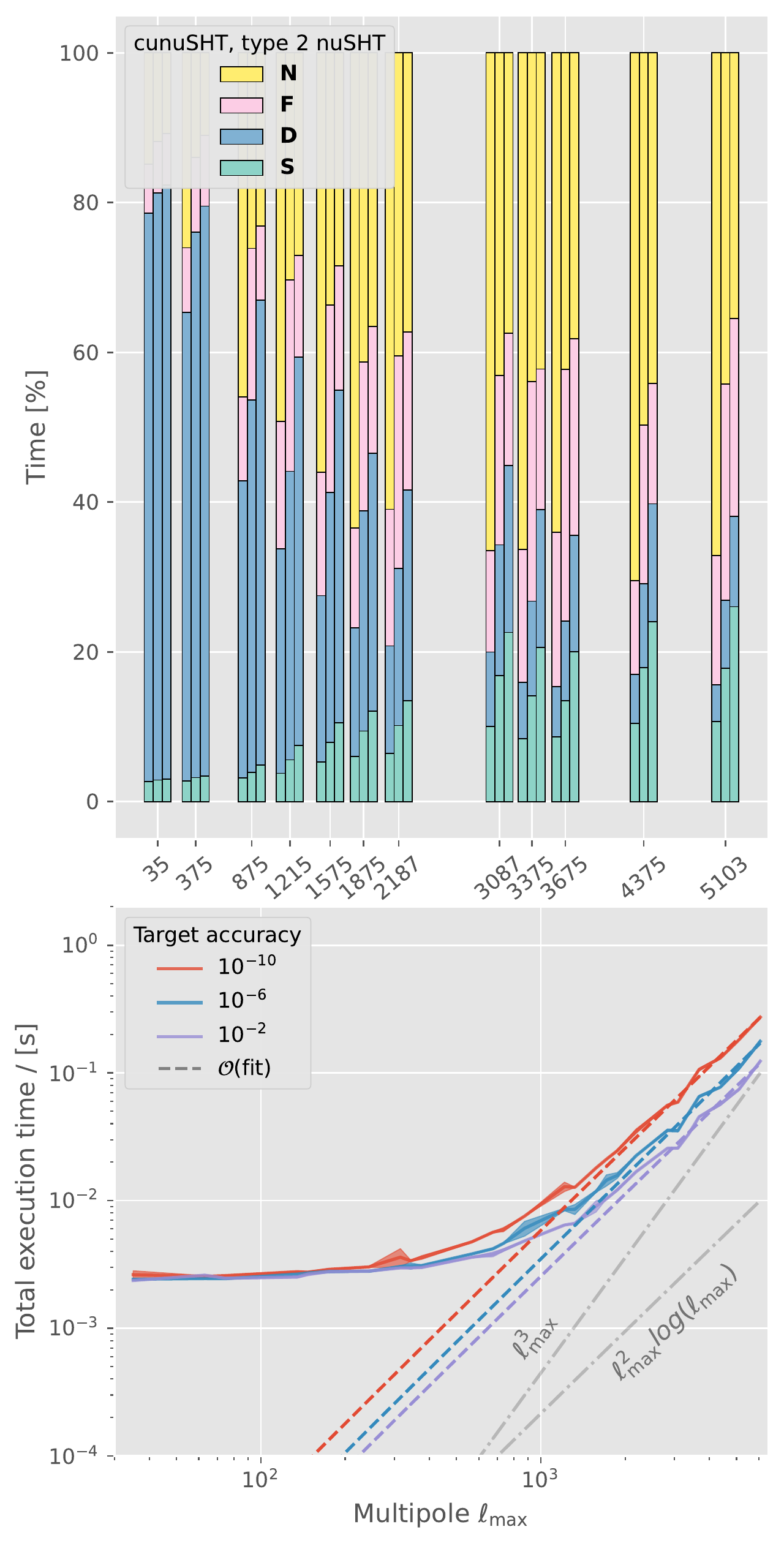}
    \includegraphics[width=0.49\textwidth]{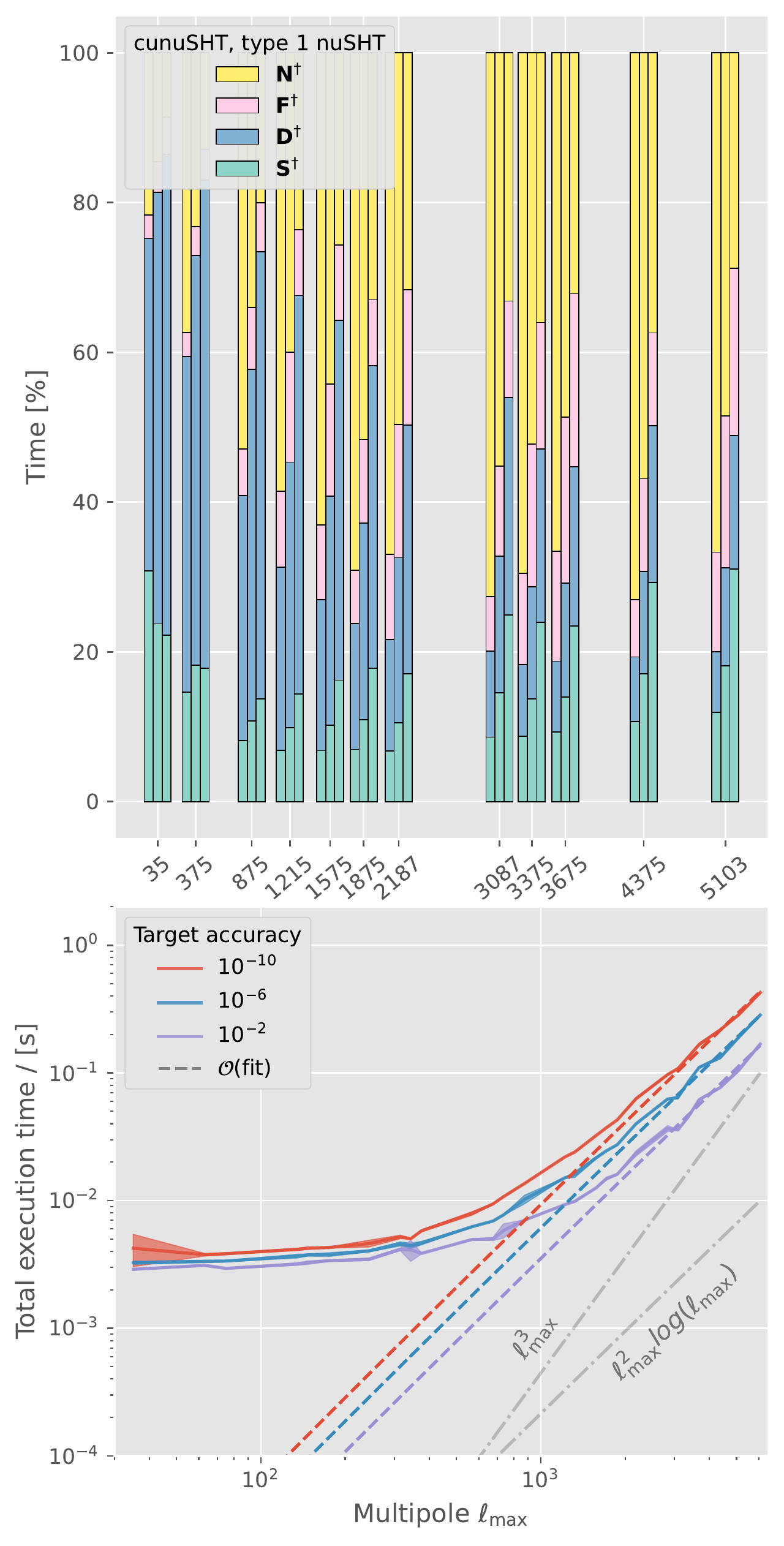}
    \caption{Breakdown and total execution time of the GPU implementation \cunusht{}. The top panels show the percentage of time spent with the individual operators for type 2 (left column) and type 1 (right column) nuSHT. For each of the problem sizes we show this for different accuracies ($10^{-10}$ on the left, $10^{-6}$ in the center, $10^{-2}$ on the right). The bottom panels show the total execution time, with $\pm1\sigma$ variance as shaded area, for different accuracies and the empirically fitted computational complexity model, Eq.~\eqref{apeq:ccx}.}
    \label{fig:GPU_synthesis_general}
\end{figure*}

\begin{figure*}
    \centering
    \includegraphics[width=0.49\textwidth]{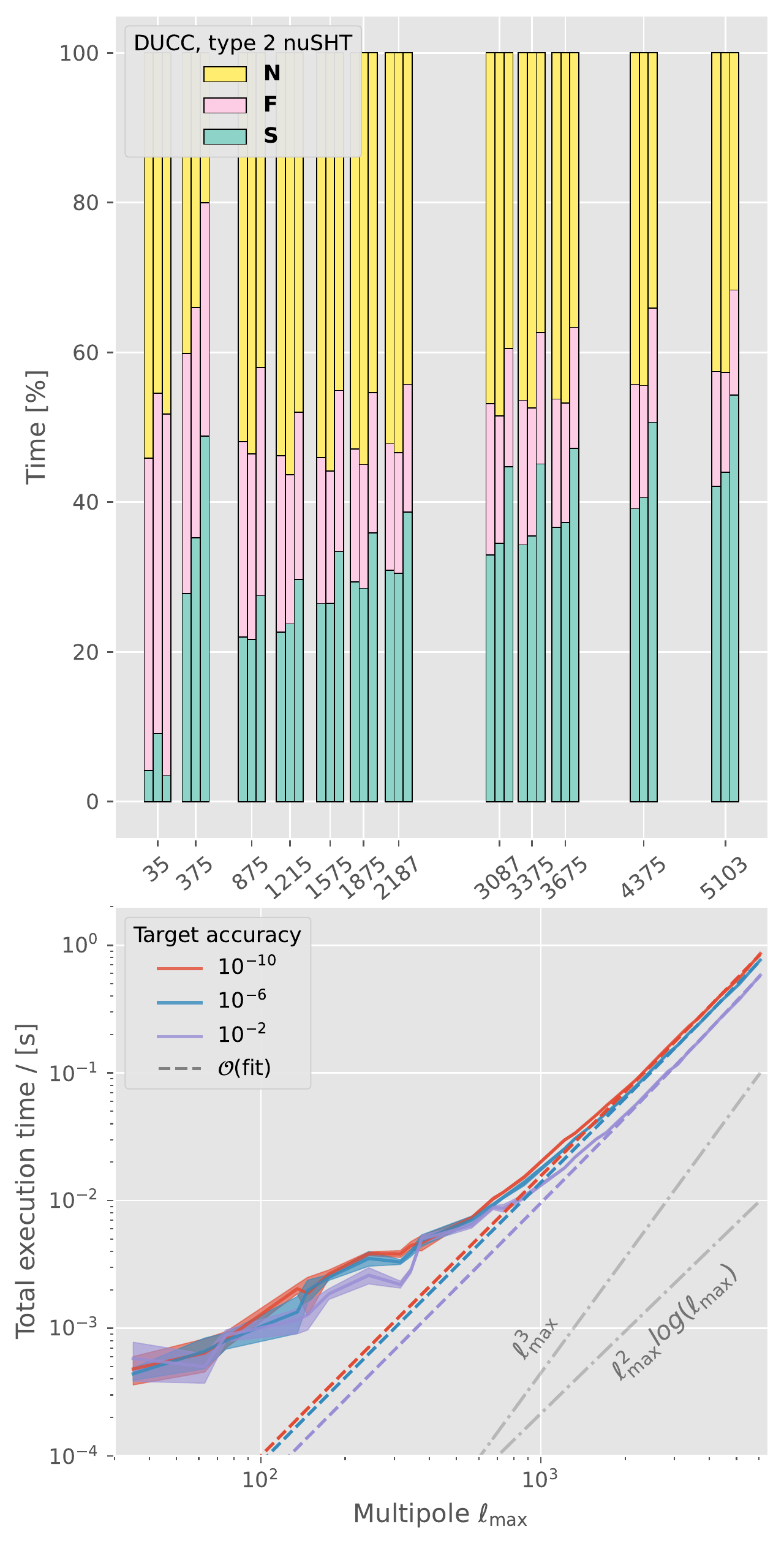}
    \includegraphics[width=0.49\textwidth]{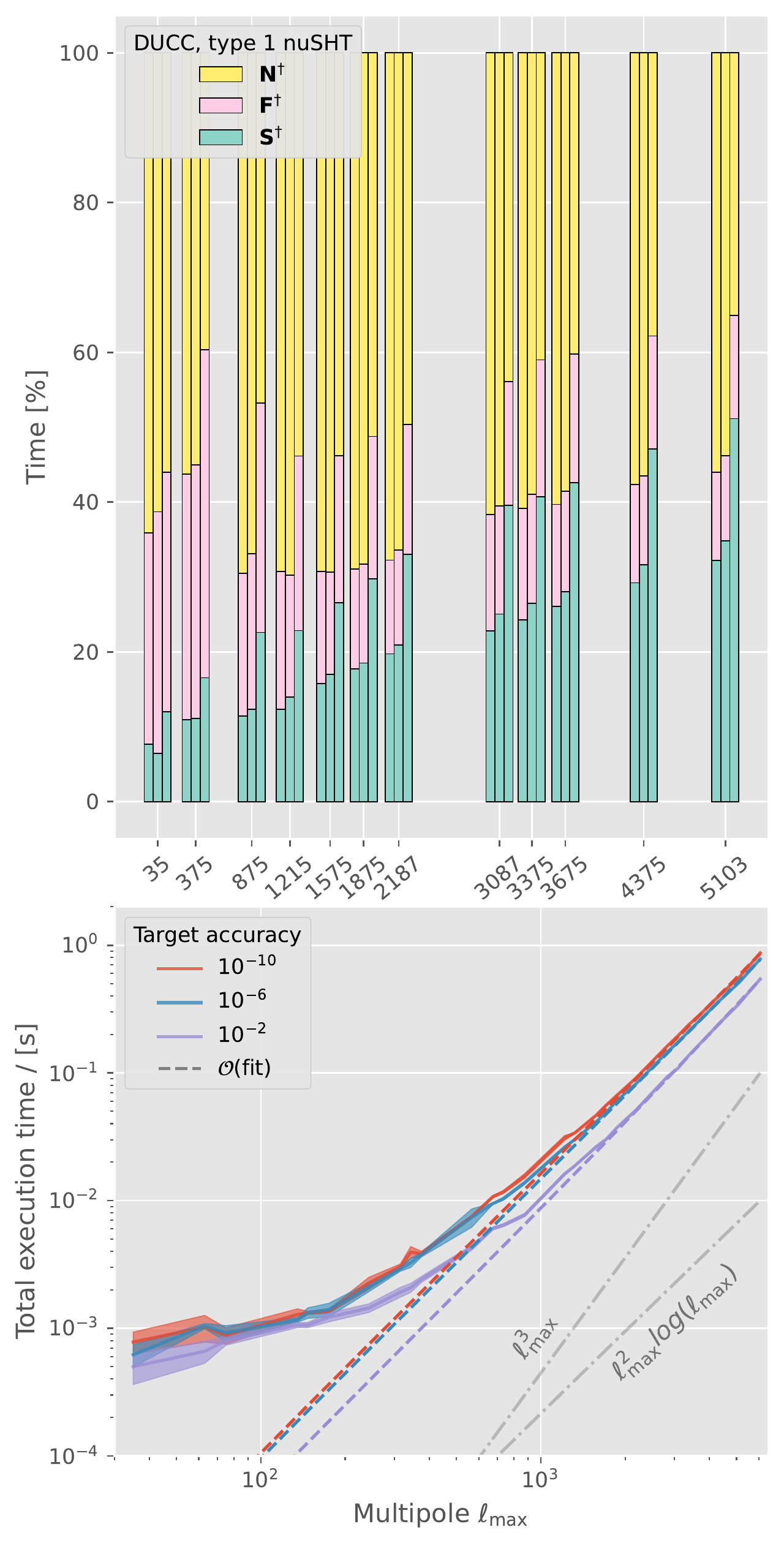}
    \caption{Same as Fig.~\ref{fig:GPU_synthesis_general}, but for the CPU implementation \ducc{}. It is important to note that \ducc{} implements the individual operators more efficiently by avoiding redundant Fourier transforms and by effectively combining the doubling and Fourier transform operations into one. To reflect this, we have grouped the $\mathbf{D}$ and $\mathbf{F}$ contributions in the top panel.}
    \label{fig:CPU_synthesis_general}
\end{figure*}
We can get a better understanding of the resulting speed up factors by looking at the time spent on each of the operators on the GPU.
This breakdown is shown in the top panels of Fig.~\ref{fig:GPU_synthesis_general} for the type 2 (left panel) and type 1 (right panel) nuSHT, as a function of $\ellmax$ for different accuracies. The respective total execution times are shown in the bottom panels together with the empirically fitted computational complexity model, Eq.~\eqref{apeq:ccx}. At the top panels, for each problem size $\ellmax$, each bar represents a benchmark with an accuracy of (from left to right) $10^{-10}$, $10^{-6}$, $10^{-2}$. Each bar represent the mean over 5 runs. For small problem sizes, doubling dominates and becomes almost negligible for large $\ellmax$. $\mathbf{S}$ only takes about 20\% of the execution time for large $\ellmax$, even though it has the worst asymptotic computational complexity. This highlights the quality of the rSHT implementation by \shtns{}. 

The choice of accuracy has an impact on the total execution time, as seen in the bottom panels of Fig.~\ref{fig:GPU_synthesis_general}. The highest accuracy ($\epsilon=10^{-10}$, red line) takes at most twice as long compared to the low accuracy ($\epsilon=10^{-2}$, purple line).
Our results suggests that improvements in $\mathbf{F}$ might be possible: due to FFT being in principle a bandwidth limited routine, we would expect the execution time of $\mathbf{F}$ to be comparable to that of $\mathbf{D}$, and only a fraction of that of $\mathbf{S}$.
For type 1 (right column), the total execution time overall takes longer. The breakdown shows that for the low and intermediate accuracy cases, less time is spend in the $\mathbf{F^{\dagger}}$ call. This is a consequence of our use of single precision arithmetic FFTs for the $\epsilon\leq 10^{-6}$ case, which, as mentioned before, is possible for the type 1 nuSHT.

It is interesting to compare this breakdown to the CPU implementation of \ducc{}. While we cannot expect both breakdowns to be exactly the same due to the different nature of the hardware, large differences may imply possible improvements. Fig.~\ref{fig:CPU_synthesis_general} shows the computation time of the CPU implementation of \ducc{} for the type 2 nuSHT (left column) and type 1 nuSHT (right column) as a function of $\ellmax$ for different accuracies. Looking at the top left panel, we see that $\mathbf{S}$ increases with increasing $\ellmax$, as expected from its $\bigO(\ellmax^3)$ scaling. The bottom panel of the left figure shows that the total execution time of the high-accuracy run is at most twice as long as the low accuracy one.

Many optimizations have gone into \ducc{}'s and \shtns{}'s SHT routines ($\mathbf{S}$, $\mathbf{S}^{\dagger}$) over the years and it is safe to assume that they are close to optimal. Comparing the breakdown of the CPU and GPU implementation for $\mathbf{S}$ (green bars) shows that \ducc{} spends more than twice as long with this operator. This hints to potential sub-optimalities in the implementations of the GPU operators. This also becomes apparent when we look at the operator $\mathbf{N}$ (and $\mathbf{N}^{\dagger}$). The scaling with the problem size is much more pronounced for the GPU code.

\begin{figure}
    \centering
    \includegraphics[width=0.49\textwidth]{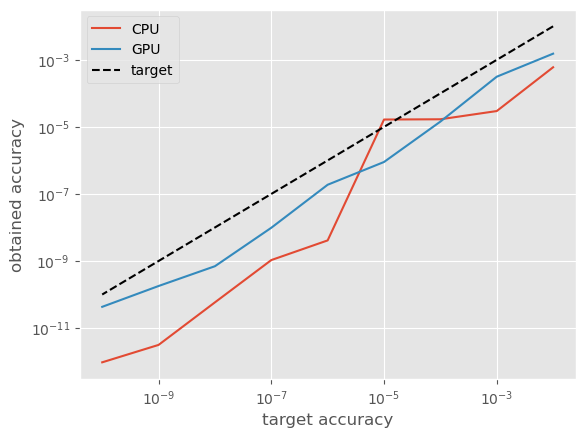}
    \caption{Effective accuracy as a function of target accuracy for the CPU (blue), and GPU (orange) implementation.}
    \label{fig:synthesis_general_accuracy}
\end{figure}

Fig.~\ref{fig:synthesis_general_accuracy} shows the effective accuracy as a function of target accuracy for both CPU and GPU. 
The effective accuracy $\epsilon_{\rm eff}$ is calculated by solving Eq.~\eqref{eq:SHTbase} in a brute-force manner (giving us the values $f^{\rm true}$) and comparing it against Eq.~\eqref{eq:SHTfast} (giving us the values $\hat{f}^{\rm est}$),
\begin{equation}
    \epsilon_{\rm eff} = \frac{1}{\bar f^{\rm true}}\sqrt{\sum_{i=1}^{N_t}\left(f_i^{\text {true}}-\hat{f}_i^{\rm est}\right)^2}\,,
\end{equation}
with
\begin{equation}
 \bar f^{\rm true}=   \sqrt{\sum_{i=1}^{N_t}\left(f_i^{\text{true}}\right)^2}.
\end{equation}
We choose $N_t=10^6$ random points on the sphere, giving us good sampling across the full sphere, and a sufficiently low variance on $\epsilon_{\rm eff}$.
Both algorithms achieve good effective accuracies, with the CPU code being more conservative. For the GPU implementation,
we note that only low effective accuracies, $\epsilon_{\rm eff} > 10^{-4}$, can be achieved when nuFFT is executed in single precision. When an accuracy of $\epsilon = 10^{-6}$ is desired, a single precision nuFFT evaluation is thus insufficient. We find that executing $\mathbf{N}$ and $\mathbf{N}^\dagger$ in double precision solves this.
There is an associated penalty in efficiency due to the double precision calculations being approximately twice as slow. The nuFFT implementation in \ducc{} circumvents this by allowing for double precision accuracy on the pointing and intermediate results while using single precision accuracy on the Fourier coefficients, hereby reducing the execution time for single precision accuracies.

\section{Conclusion}\label{sec:conclusion}
We presented \cunusht{}, a GPU accelerated implementation of the spherical harmonic transform on arbitrary pixelization that is, to our knowledge, the first of its kind to achieve faster execution when compared against CPU based algorithms. \cunusht{} achieves machine precision accuracy by transforming the problem of interpolating on the  sphere into a problem of computing a nonuniform fast Fourier transform on the torus.
Comparing our implementation executed on an \GPUdesc{} to the fastest available CPU implementations to date running on a single \CPUdesc{}  with \nthreads{} cores, we find that our implementation is up to \suf{} times faster. We used highly efficient, publicly available packages that are well tested and robust: \shtns{} for rSHTs, \cufinufft{} for nuFFTs. We found that, although it has the highest asymptotic complexity, the high-quality rSHT implemented in SHTns is not the bottleneck. Our code does not require intermediate transfers between host and device, allowing it to be incorporated within larger GPU-based algorithms. Many applications in cosmology that we have in mind typically require spin-$1$ to $3$ transforms. 
We thus plan to extend this package to spin-$n$ transforms in the near future. There are, in principle, no obstacles to the generalization to spin-$n$ by implementing the corresponding Wigner-$d$ transform. 

\cunusht{} is a general purpose package distributed via pypi, and also works on standard pixelization schemes such as \texttt{HEALPix}, and can also perform rSHTs. Demonstrations of our package on GitHub present how to integrate it into exisiting pipelines.

\section*{Acknowledgment}
The authors thank Alex Barnett for helpful discussions about nuFFT, Lehman Garrison for help on the cupy-nanobind implementation, and Libin Lu for general discussion and upgrades to \cufinufft{}. This work is supported by the Research Analyst grant from the Simons Foundation and the computing resources of the Flatiron Institute. The Flatiron Institute is supported by the Simons Foundation. SB and JC acknowledges support from a SNSF Eccellenza Professorial Fellowship (No. 186879).

\bibliography{references}

\renewcommand\theequation{\Alph{section}\arabic{equation}} 
\counterwithin*{equation}{section} 
\renewcommand\thefigure{\Alph{section}\arabic{figure}}
\counterwithin*{figure}{section}
\renewcommand\thetable{\Alph{section}\arabic{table}}
\counterwithin*{table}{section}
\appendix

\section{Data Transfer}\label{ap:transfer}
\begin{figure}
    \centering
    \includegraphics[width=0.49\textwidth]{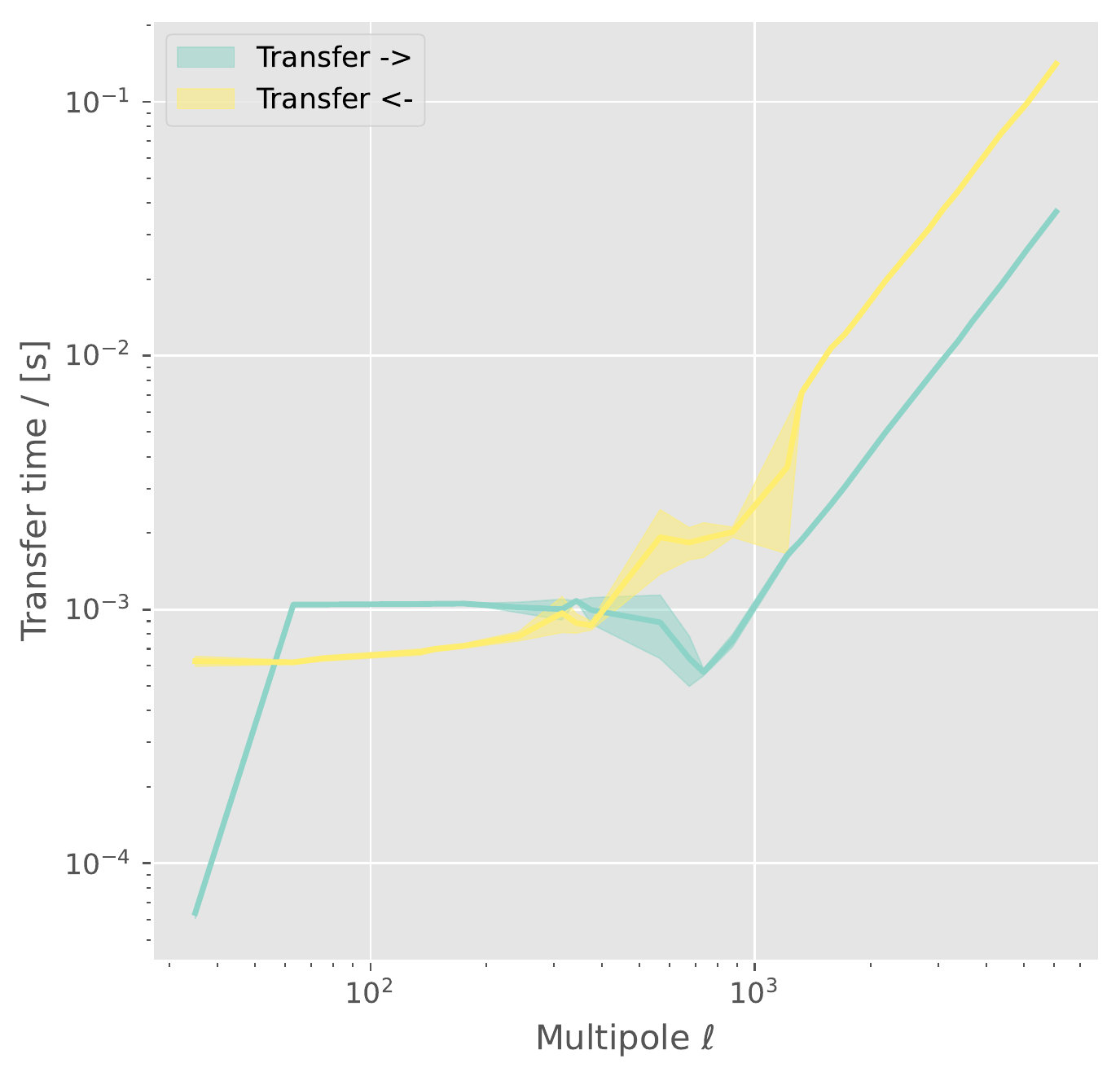}
    \caption{type 2 nuSHT transfer times ``Host-to-Device'' (H2D) (green) and ``Device-to-Host'' (D2H) (yellow) as a function of $\ellmax$. The shaded areas show the $\pm 1\sigma$ variance in transfer time, calculated from ten runs. Note that these transfer times are not generally part of the routine and can completely be avoided. H2D transfer contains the SHT coefficients, while D2H contains the much larger map. For type 1 nuSHT, the transfer times are equally long, but the H2D and D2H are swapped due to the input and output of the function.}
    \label{fig:transfer}
\end{figure}

All benchmarks are done without measuring the time to transfer the data to device and back. This can be a large part of the overall computation and should be avoided, as is shown in Fig.~\ref{fig:transfer} for different problem sizes. We note here that \cunusht{} provides means to keep everything on the GPU without having to transfer the data. Thus, these transfer times can in principle be avoided.

\section{Good numbers}\label{ap:goodnumber}
For fast Fourier transforms, execution time depends strongly on the largest prime factor of the transform length; the smaller it is, the better. Luckily enough, many good numbers for CPU algorithms exist. Assuming Clenshaw-Curtis quadrature, the constraint on the number of rings $N_r$ to sufficiently sample a map with band limit $\ellmax$ is $N_r \geq \ellmax+1$. Therefore, after applying the double Fourier sphere method, we must have at least $2N_r+2$ samples, and we are free to choose an arbitrarily larger number which happens to be a good FFT size. For the GPU algorithm, however, there is a caveat: due to the additional constraint on $N_r$ to be divisible by $4$ for the \rsht{} operation, the selection of good numbers is reduced. While there are still good numbers for FFT satisfying both constraints, they are sparser. The full list of $N_r$ that are multiples of 4 up to $N_r\leq 10000$ for which $\ellmax = N_r-1$ can be factored into primes up to 11 is 
4, 8, \underline{12}, 16, 28, 36, \underline{56}, 64, 76, \underline{100}, 136, 148, 176, \underline{232}, 244, \underline{276}, 316, 344, \underline{364}, 376, \underline{496}, \underline{540}, 568, 676, 736, \underline{848}, 876, \underline{892}, \underline{1156}, 1216, 1324, \underline{1332}, \underline{1376}, 1576, 1716, \underline{1816}, 1876, \underline{2080}, 2188, \underline{2476}, \underline{2696}, 2836, 3088, \underline{3268}, 3376, 3676, \underline{4236}, 4376, \underline{4456}, \underline{4852}, 5104, \underline{5776}, 6076, 6616, \underline{6656}, \underline{6876}, 7204, \underline{7624}, 7876, \underline{8020}, 8576, \underline{9076}, 9376.
The numbers with the highest prime factor being 11 are highlighted with an underline as efficient calculation of \cunusht{} only applies to the CPU code here.

One way to enlarge this list could be for example by using a Fejér grid as intermediate grid, as opposed to the currently used Clenshaw-Curtis grid. This would effectively reduce the number of required samples to $2N_r$ after doubling, resulting in a much wider choice of sample sizes that are also good FFT lengths. Hence restrictions on the number of good numbers are expected to go away in the future.

\section{CMB weak lensing pointing}\label{ap:pointing}
\begin{figure}
    \centering
    \includegraphics[width=0.49\textwidth]{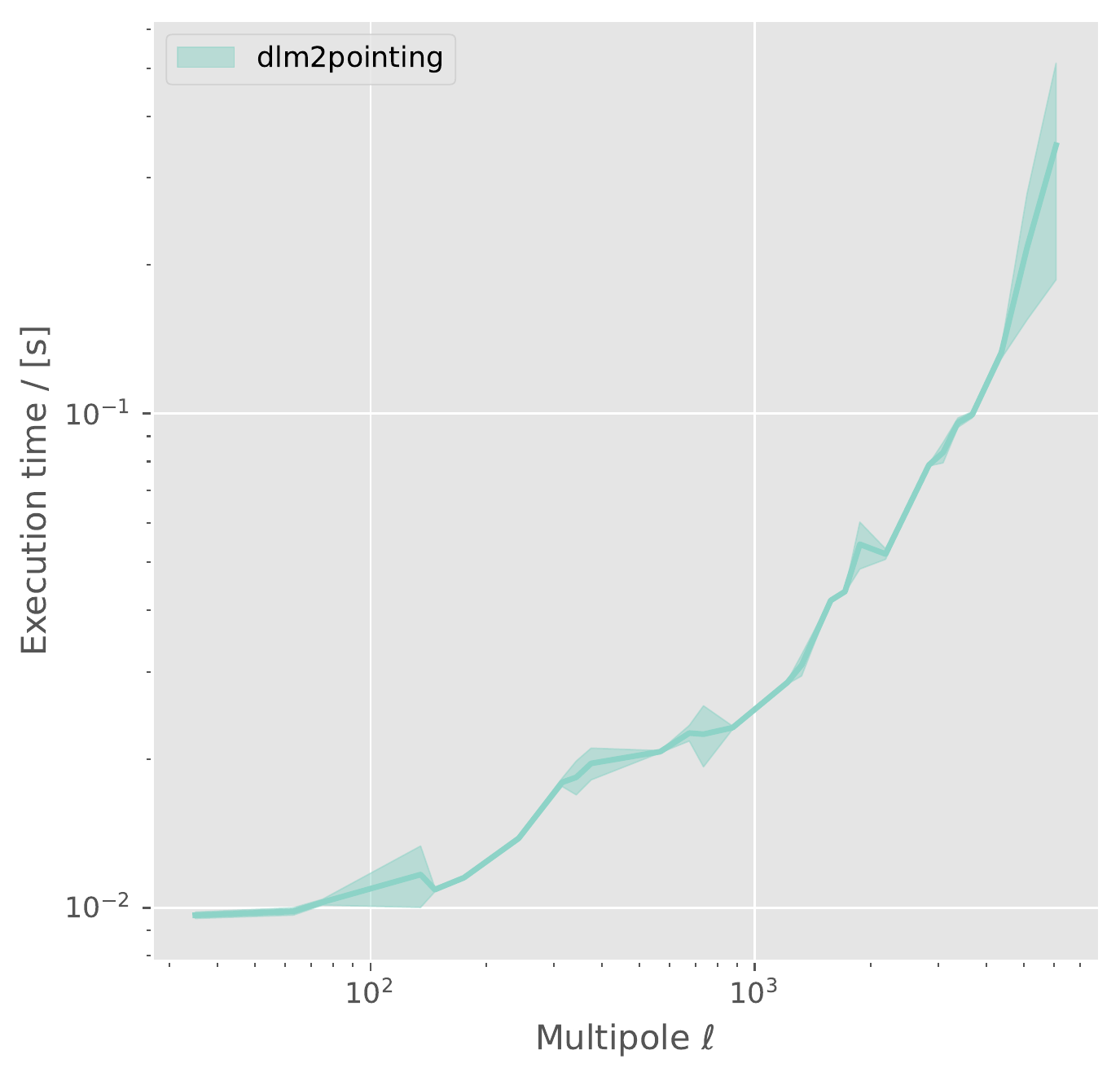}
    \caption{Execution time of the pointing routine on the GPU as a function of the problem size. The shaded area shows the $\pm1\sigma$ uncertainty calculated from 5 runs.}
    \label{fig:pointing}
\end{figure}

Let $\hat{e}_\theta, \hat{e}_\phi$ and $\hat n$ form the right-handed unit basis vectors at the point on the sphere parametrized by $\theta, \phi$ (so, the components of $ \hat n$ are $\sin \theta \cos \phi, \sin \theta \sin \phi, \cos \theta$). In our benchmark CMB lensing application, the deflected positions $\hat n'$ that define the angles $\theta'$ and $\phi'$ at which the CMB field must be evaluated are given by,
\begin{equation}\label{eq:pointing}
    \hat n' = \cos(\alpha)\hat{n}+\frac{\sin(\alpha)}{\alpha}\left(\alpha_\theta\hat{e}_\theta+\alpha_\phi\hat{e}_\phi\right)\,,
\end{equation}
where $\alpha = \sqrt{\alpha_\theta^2 + \alpha_\phi^2}$, and $\alpha_\theta$ and $\alpha_\phi$ form the gradient of the lensing potential $\Phi$,
\begin{equation}\label{eq:spin1deriv}
    \alpha_\theta(\hat n) + \iu\alpha_\phi(\hat n) = \left(\frac{\partial }{\partial \theta} + \frac{\iu}{\sin \theta} \frac{\partial }{\partial \phi} \right) \Phi(\hat n).
\end{equation}
\cunusht{} first obtains $\alpha_\theta$ and $\alpha_\phi$ from this equation using \shtns. Then, threading across each ring, we solve for $\theta', \phi'$ in~Eq.~\eqref{eq:pointing} on the fly. A benchmark as a function of $\ellmax$ is shown in Fig.~\ref{fig:pointing}.

\section{Computational complexity}\label{ap:ccx}
Table \ref{tab:t2} shows the results for the empirical fit of the type 2 and type 1 nuSHT computational complexity models,
\begin{equation}\label{apeq:ccx}
    C(\ell) = \alpha\frac{\ell^3}{\ell_{\rm norm}^3} + \beta\frac{\ell^2\log(\ell)}{\ell_{\rm norm}^2\log(\ell_{\rm norm})} \,,
\end{equation}
with $\ell_{\rm norm} = 6067$ the normalization for the unknown prefactors. Compared to Eq.~\eqref{eq:ccx} we only account for the terms with the largest complexities. The fits are shown in Fig.~\ref{fig:speedup}, Fig.~\ref{fig:GPU_synthesis_general}, and Fig.~\ref{fig:CPU_synthesis_general}.

\begin{table}[]
\caption{Empirical fits of Eq.~\eqref{eq:ccx} for type 2 and type 1 nuSHT, for different accuracies for both CPU and GPU.
}\label{tab:t2}
\begin{tabular}{llcc}
\toprule
type 2\ &\ Target acc.\ &\ \ \ \ \ \ \ $\alpha$\ \ \ \ \ \ \ \ &\ \ \ \ \ \ \ $\beta$\ \ \ \ \ \ \ \\
\midrule

\multirow{3}{*}{CPU} & $10^{-10}$ & $1.52\cdot 10^{-1}$ & $6.96\cdot 10^{-1}$ \\
                     & $10^{-6}$  & $1.41\cdot 10^{-1}$ & $6.16\cdot 10^{-1}$ \\
                     & $10^{-2}$  & $1.67\cdot 10^{-1}$ & $4.11\cdot 10^{-1}$ \\
\midrule
\multirow{3}{*}{GPU} & $10^{-10}$ & $1.04\cdot 10^{-6}$ & $2.72\cdot 10^{-1}$ \\
                     & $10^{-6}$  & $1.06\cdot 10^{-2}$ & $1.60\cdot 10^{-1}$ \\
                     & $10^{-2}$  & $1.39\cdot 10^{-4}$ & $1.19\cdot 10^{-1}$ \\
\toprule
type 1\ &\ Target acc.\ & $\alpha$ & $\beta$ \\
\midrule
\multirow{3}{*}{CPU} & $10^{-10}$ & $1.24\cdot10^{-1}$ & $7.34\cdot 10^{-1}$ \\
                     & $10^{-6}$  & $1.06\cdot10^{-1}$ & $6.67\cdot 10^{-1}$ \\
                     & $10^{-2}$  & $1.68\cdot10^{-1}$ & $3.72\cdot 10^{-1}$ \\
\midrule
\multirow{3}{*}{GPU} & $10^{-10}$ & $1.31\cdot10^{-8}$ & $4.33\cdot 10^{-1}$ \\
                     & $10^{-6}$  & $1.07\cdot10^{-6}$ & $2.84\cdot 10^{-1}$ \\
                     & $10^{-2}$  & $2.16\cdot10^{-5}$ & $1.64\cdot 10^{-1}$ \\
\bottomrule
\end{tabular}
\end{table}

\section{Code examples}
The following code block shows a minimum working example for calculating a type 2 and type 1 nuSHT on the GPU for SHT coefficients \texttt{alm} that are deflected by a deflection field \texttt{dlm\_scaled}, for an accuracy of \texttt{epsilon} and band limit \texttt{lmax}. Both routines will set up the plans, so that the actual \texttt{nusht2dX()} call may be called repeatedly.
\begin{minipage}{\linewidth}
\begin{lstlisting}[language=Python]
import cunusht as cu

lenjob_geominfo = ('gl',{'lmax': lmax})
kwargs = {
    'geominfo_deflection': lenjob_geominfo,
    'epsilon': epsilon,
    'nuFFTtype': 2,
}
t = cu.get_transformer(backend='GPU')(**kwargs)
ptg = t.dlm2pointing(dlm_scaled)
lenmap = t.nusht2d2(alm, ptg, lmax, lenmap)
\end{lstlisting}
\end{minipage}
\begin{minipage}{\linewidth}
\begin{lstlisting}[language=Python]
import cunusht as cu

lenjob_geominfo = ('gl',{'lmax': lmax})
kwargs = {
    'geominfo_deflection': lenjob_geominfo,
    'epsilon': epsilon,
    'nuFFTtype': 1,
}
t = cu.get_transformer(backend='GPU')(**kwargs)
ptg = t.dlm2pointing(dlm_scaled)
alm = t.nusht2d1(alm, ptg, lmax, lenmap)
\end{lstlisting}
\end{minipage}
\end{document}